\documentclass[11pt]{article}

\usepackage{amsmath,amsthm,amssymb}
\usepackage{graphicx}
\usepackage{url}

\DeclareMathOperator{\var}{var}

\begin{document}

\title{Novelty and Collective Attention}
\author{Fang Wu and Bernardo A. Huberman\\ \small{Information Dynamics Laboratory}\\ \small{HP Labs}\\ \small{Palo Alto, CA 94304}}

\maketitle

\begin{abstract}
The subject of collective attention is central to an information age
where millions of people are inundated with daily messages. It is
thus of interest to understand how attention to novel items
propagates and eventually fades among large populations. We have
analyzed the dynamics of collective attention among one million
users of an interactive website --- \texttt{digg.com} --- devoted to
thousands of novel news stories. The observations can be described
by a dynamical model characterized by a single novelty factor. Our
measurements indicate that novelty within groups decays with a
stretched-exponential law, suggesting the existence of a natural
time scale over which attention fades.
\end{abstract}

\pagebreak

The problem of collective attention is at the heart of decision
making and the spread of ideas, and as such it has been studied at
the individual and small group level by a number of psychologists
\cite{kahneman-73, pashler-98}, economists \cite{camerer-03}, and
researchers in the area of marketing and advertising \cite{PRW-99,
dukas-04, reis-06}. Attention also affects the propagation of
information in social networks, determining the effectiveness of
advertising and viral marketing \cite{LAH-06}. And while progress on
this problem has been made in small laboratory studies and in the
theoretical literature of attention economics \cite{falkinger-03},
it is still lacking empirical results from very large groups in a
natural, non-laboratory, setting.

To understand the process underlying attention in large groups,
consider as an example how a news story spreads among a group of
people. When it first comes out, the story catches the attention of
a few ones, who may further pass it on to others if they find it
interesting enough. If a lot of people start to pay attention to
this story, its exposure in the media will continue to increase. In
other words, a positive reinforcement effect sets in such that the
more popular the story becomes, the faster it spreads.

This growth is counterbalanced by the fact that the novelty of a
story tends to fade with time and thus the attention that people pay
to it. Therefore, in considering the dynamics of collective
attention two competing effects are present: the growth in the
number of people that attend to a given story and the habituation
that makes the same story less likely to be attractive as time goes
on. This process becomes more complex in the realistic case of
multiple items or stories appearing at the same time, for now people
also have the choice of which stories to attend with their limited
attention.

In order to study the dynamics of collective attention and its
relation to novel inputs in a natural setting, we analyzed the
behavioral patterns of one million people interacting with a news
website whose content is solely determined by its own users. Because
people using this website assign each news story an explicit measure
of popularity, we were able to determine the growth and decay of
attention for thousands of news stories and to validate a
theoretical model which predicts both the dynamics and the
statistical distribution of story lifetimes.

The website under study, \url{digg.com}, is a digital media
democracy which allows its users to submit news stories they
discover from the Internet \cite{digg.com}. A new submission
immediately appears on a repository webpage called ``Upcoming
Stories'', where other members can find the story and, if they like
it, add a \emph{digg} to it. A so-called \emph{digg number} is shown
next to each story's headline, which simply counts how many users
have \emph{digged} the story in the past.\footnote{In fact, digg
users are given the option to ``bury'' a story, which will
\emph{decrease} the story's digg number. Because this rarely
happens, we ignore this possibility and simply assume that a story's
digg number can only grow with time.} If a submission fails to
receive enough diggs within a certain time limit, it eventually
falls out of the ``Upcoming'' section, but if it does earn a
critical mass of diggs quickly enough, it becomes \emph{popular} and
jumps to the \url{digg.com} front page.\footnote{The actual
machine-learning algorithm used to determine whether a story
qualifies to appear on the front page is very complex and will not
be discussed in this paper \cite{digg-algorithm}.} Because the front
page can only display a limited number of stories, old stories
eventually get replaced by newer stories as the page gets constantly
updated. If a story however, becomes very popular it qualifies as a
``Top 10'' and stays on the right side of the front page for a very
long time.

%

\begin{figure}
\centering
\begin{minipage}{2.25in}
\centering
\includegraphics[width=2.25in]{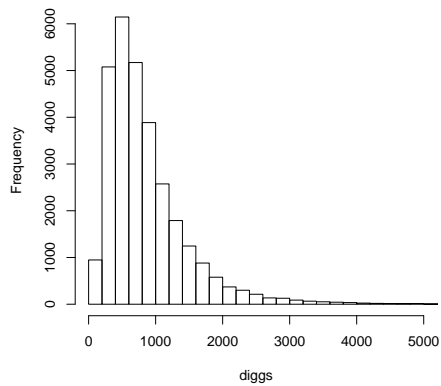}\\ \small{(a)}
\end{minipage}
\hfill
\begin{minipage}{2.25in}
\centering
\includegraphics[width=2.25in]{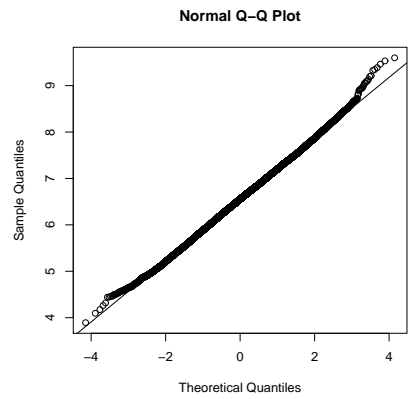}
\\ \small{(b)}
\end{minipage}
\caption{\label{fig:2006} (a) The histogram of the 29,684 diggs in
2006, as on January 9, 2007. (b) The normal Q-Q plot of
$\log(N_\infty)$. The straight line shows that $\log(N_\infty)$
follows a normal distribution with a slightly longer tail. This is
due to \texttt{digg.com}'s built-in reinforcement mechanism that
favors those ``top stories'', which can stay on the front page and
can be found at many other places (e.g.~``popular stories in 30
days'' and ``popular stories in 365 days'').}
\end{figure}

When a story first appears on the front page it attracts much
attention, causing its digg number, $N_t$, to build up quickly.
After a couple of hours its digg rate slows down because of both its
lack of novelty and its lack of prominent visibility (reflected in
the fact that it moves away from the front page). Thus the digg
number of each story eventually saturates to a value, $N_\infty$,
that depends on both its popularity growth and its novelty decay. In
order to determine the statistical distribution of this saturation
number, which corresponds to the number of diggs it has accumulated
throughout its evolution, we measured the histogram of the final
diggs of all 29,864 popular stories in the year 2006. As can be seen
from Fig.~\ref{fig:2006}, the distribution appears to be quite
skewed, with the normal Q-Q plot of $\log(N_\infty)$ a straight
line. A Kolmogorov-Smirnov normality test of $\log(N_\infty)$ with
mean 6.546 and standard deviation 0.6626 yields a $p$-value of
0.0939, suggesting that $N_\infty$ follows a log-normal
distribution.

It is then natural to ask whether $N_t$, the number of diggs of a
popular story after finite time $t$, also follows a log-normal
distribution. To answer this question, we tracked the digg numbers
of 1,110 stories in January 2006 minute by minute. The distribution
of $\log(N_t)$ again obeys a bell shape curve. As an example, a
Kolmogorov-Smirnov normality test of $\log(N_{\mathrm{2\,hours}})$
with mean 5.925 and standard deviation 0.5451 yields a $p$-value as
high as 0.5605, supporting the hypothesis that $N_t$ also follows a
log-normal distribution.


The log-normal distribution can be explained by a simple stochastic
dynamical model which we now describe. If $N_t$ represents the
number of people who know the story at time $t$, in the absence of
any habituation, \emph{on average} a fraction $\mu$ of those people
will further spread the story to some of their friends.
Mathematically this assumption can be expressed as $N_t = (1+X_t)
N_{t-1}$, where $X_1, X_2, \dots$ are positive i.i.d.~random
variables with mean $\mu$ and variance $\sigma^2$. The requirement
that $X_i$ must be positive ensures that $N_t$ can only grow with
time. As we have discussed above, this growth in time is eventually
curtailed by a decay in novelty, which we parameterize by a time
dependent \emph{factor} $r_t$ consisting of a series of decreasing
positive numbers with the property that $r_1=1$ and $r_t\downarrow
0$ as $t\uparrow\infty$. With this additional parameter, the full
stochastic dynamics of story propagation is governed by $N_t = (1+
r_t X_t) N_{t-1}$, where the factor $r_t X_t$ acts as a discounted
random multiplicative factor. When $X_t$ is small (which is the case
for small time steps) we have the following approximate solution:
\begin{equation}
N_t = \prod_{s=1}^t (1+ r_s X_s) N_0 \approx \prod_{s=1}^t e^{r_s
X_s} N_0 = e^{\sum_{s=1}^t r_s X_s} N_0,
\end{equation}
where $N_0$ is the initial population that is aware of the story.
Taking logarithm of both sides, we obtain
\begin{equation}
\label{eq:logNt}\log N_t - \log N_0 = \sum_{s=1}^t r_s X_s.
\end{equation}
The right hand side is a discounted sum of random variables, which
for $r_t$ near one (small time steps) can be shown to be described
by a normal distribution \cite{embrechts-maejima-84}. It then
follows that for large $t$ the probability distribution of $N_t$
will be approximately log-normal.

\begin{figure}
\centering
\includegraphics[scale=.5]{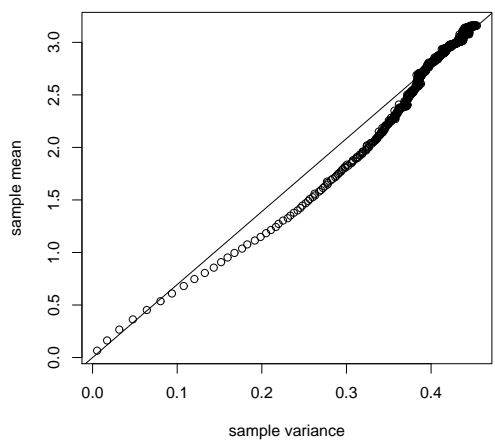}
\caption{\label{fig:mean_variance}Sample mean of $\log N_t-\log N_0$
versus sample variance, for 1,110 stories in January 2007. Time unit
is one minute. The points are plotted as follows. For each story we
calculate the quantity $\log N_t-\log N_0$, which is the logarithm
of its digg number measured $t$ minutes after its first appearance
on the front page, minus the logarithm of its initial digg number.
We collect 1,110 such quantities for 1,110 stories. Then we compute
their sample mean $y$ and sample variance $x$, and mark the point
$(x,y)$. This is for one $t$. We repeat the process for
$t=1,2,\dots, 1440$ and plot 1440 points in total (i.e.~24 hours).
They lie roughly on a straight line passing through the origin with
slope 6.947.}
\end{figure}

Our dynamic model can be further tested by taking the mean and
variance of both sides of Eq.~(\ref{eq:logNt}):
\begin{equation}
\frac{E(\log N_t - \log N_0)}{\var(\log N_t - \log N_0)} =
\frac{\sum_{s=1}^t r_s \mu} {\sum_{s=1}^t r_s \sigma^2} =
\frac{\mu}{\sigma^2}.
\end{equation}
Hence if our model is correct, a plot of the sample mean of
$\log(N_t)-\log(N_0)$ versus the sample variance for each time $t$,
should yield a straight line passing through the origin with slope
$\mu/\sigma^2$. One such plot for 1,110 stories collected in January
2007 is shown in Fig.~\ref{fig:mean_variance}. As can be seen, the
points indeed lie on a line with slope 6.947.

\begin{figure}
\centering
\begin{minipage}{2.25in}
\centering
\includegraphics[width=2.25in]{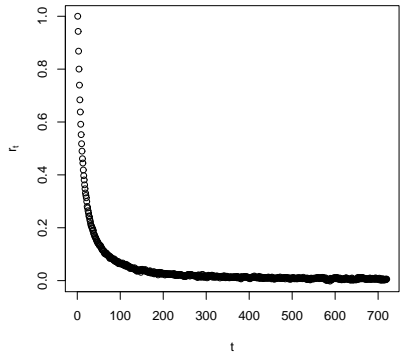}\\ \small{(a)}
\end{minipage}
\hfill
\begin{minipage}{2.25in}
\centering
\includegraphics[width=2.25in]{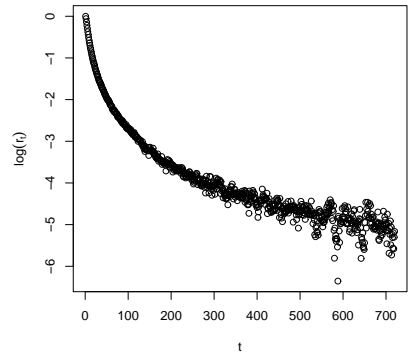}\\ \small{(b)}
\end{minipage}
\begin{minipage}{2.25in}
\centering
\includegraphics[width=2.25in]{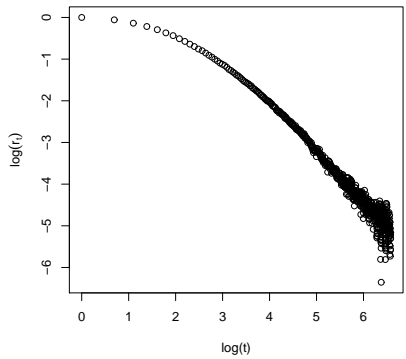}\\ \small{(c)}
\end{minipage}
\hfill
\begin{minipage}{2.25in}
\centering
\includegraphics[width=2.25in]{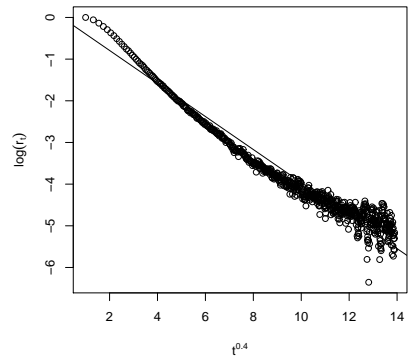}\\ \small{(d)}
\end{minipage}
\caption{\label{fig:r} (a) The decay factor $r_t$ as a function of
time. Time $t$ is measured in minutes. (b) $\log(r_t)$ versus $t$.
$r_t$ decays slower than exponential. (c) $\log(r_t)$ versus $t$.
$r_t$ decays faster than power law. (d) $\log(r_t)$ versus
$t^{0.4}$. The slope is approximately $-0.4$.}
\end{figure}

The decay factor $r_t$ can now be computed explicitly from $N_t$ up
to a constant scale. Since we have normalized $r_1$ to 1, we have
\begin{equation}
r_t = \frac{E(\log N_t)-E(\log N_{t-1})} {E(\log N_1)-E(\log N_0)}.
\end{equation}
The curve of $r_t$ estimated from the 1,110 stories in January 2007
is shown in Fig.~\ref{fig:r}(a). As can be seen, $r_t$ decays very
fast in the first two to three hours, and its value becomes less
than 0.03 after three hours. Fig.~\ref{fig:r}(b,c) show that $r_t$
decays slower than exponential and faster than power law.
Fig.~\ref{fig:r}(d) shows that $r_t$ can be fit empirically to a
\emph{stretched exponential relaxation} or Kohlrausch-Williams-Watts
law \cite{lindsey-patterson-80}: $r_t \sim e^{-0.4 t^{0.4}}$. The
\emph{halflife} $\tau$ of $r_t$ can then be determined by solving
the equation
\begin{equation}
\int_0^\tau e^{-0.4 t^{0.4}} dt = \frac 12 \int_0^\infty e^{-0.4
t^{0.4}} dt.
\end{equation}
A numerical calculation gives $\tau=69$ minutes, or about one hour.
This characteristic time is consistent with the fact that a story
usually lives on the front page for a period between one and two
hours.

The stretched exponential relaxation often occurs as the result of
multiple characteristic relaxation time scales
\cite{lindsey-patterson-80, frisch-sornette-97}. This is consistent
with the fact that the decay rate of a story on \url{digg.com}
depends on many factors, such as the story's topic category, the
time of a day when it appears on the front page. The measured decay
factor $r_t$ is thus an average of these various rates and describes
the collective decay of attention.

These measurements, comprising the dynamics of one million users
attending to thousands of novel stories, allowed us to determine the
effect of novelty on the collective attention of very large groups
of individuals, while nicely isolating both the speed of propagation
of new stories and their decay. We also showed that the growth and
decay of collective attention can be described by a dynamical model
characterized by a single novelty factor which determines the
natural time scale over which attention fades. The exact value of
the decay constant depends on the nature of the medium but its
functional form is universal. These experiments, which complement
large social network studies of viral marketing \cite{LAH-06} are
facilitated by the availability of websites that attract millions of
users, a fact that turns the internet into an interesting natural
laboratory for testing and discovering the behavioral patterns of
large populations on a very large scale \cite{watts-07}.



\begin{thebibliography}{99}

\bibitem{kahneman-73} Kahneman, D. Attention and effort. Englewood
Cliffs, N.J.: Prentice Hall (1973).

\bibitem{pashler-98} Pashler, H. E. The psychology of attention.
MIT Press (1998).

\bibitem{camerer-03} Camerer, C. The behavioral challenge to
economics: understanding normal people. Paper presented at Federal
Reserve of Boston meeting (2003).

\bibitem{PRW-99} Pieters, F. G. M., Rosbergen, E. and Wedel, M. Visual attention
to repeated print advertising: A test of scanpath theory.
\textit{Journal of Marketing Research} 36(4), 424--438 (1999).

\bibitem{dukas-04}Dukas, R. Causes and consequences of limited attention.
\textit{Brain Behavior and Evolution} 63, 197--210 (2004).

\bibitem{reis-06} Reis, R. Inattentive Consumers. \textit{Journal of Monetary
Economics} Vol.~53, 1761--1800 (2006).

\bibitem{LAH-06} Lefkovic, J., Adamic, L. and Huberman, B. A. The dynamics of
viral marketing. \textit{Proceedings of the ACM Conference on
Electronic Commerce} (2006).

\bibitem{falkinger-03} Falkinger, J. Attention Economies.
Forthcoming in \textit{Journal of Economic Theory} (2003).

\bibitem{digg.com} How Digg Works. \url{http://www.digg.com/how}

\bibitem{digg-algorithm} Private communication with the
\url{digg.com} support team.

\bibitem{embrechts-maejima-84} Embrechts, P. and Maejima, M.
The central limit theorem for summability methods of i.i.d.~random
variables. \textit{Probability Theory and Related Fields}, Volume
68, Number 2 (1984).

\bibitem{lindsey-patterson-80} Lindsey, C. P. and Patterson, G. D.
Detailed comparison of the William-Watts and Cole-Davidson
functions. \textit{J. Chem.~Phys.}~\textbf{73}(7) (1980).

\bibitem{frisch-sornette-97} Frisch, U. and Sornette, D. Extreme deviations and
applications. \textit{J. Phys.~I France} \textbf{7} 1155--1171
(1997).

\bibitem{watts-07} Watts, D. A twenty first century science. \textit{Nature} \textbf{445}, pp.~489
(2007).
\end{thebibliography}
\end{document}